\documentclass[letterpaper, 10 pt, conference]{ieeeconf}  
\IEEEoverridecommandlockouts
\makeatletter
\let\NAT@parse\undefined
\makeatother

\usepackage[dvipsnames]{xcolor}

\usepackage{times}
\usepackage{graphicx}
\usepackage{amssymb}
\usepackage{gensymb}
\usepackage{amsmath}
\usepackage{breakurl}

\usepackage{url,hyperref}

\usepackage{algorithm}
\usepackage{algorithmic}

\usepackage[labelfont={bf},font=small]{caption}
\usepackage[none]{hyphenat}

\usepackage{mathtools, cuted}

\usepackage[noadjust, nobreak]{cite}

\usepackage{tabularx}
\usepackage{amsmath}

\usepackage{float}

\usepackage{pifont}

\newcolumntype{Y}{>{\centering\arraybackslash}X}

\usepackage[]{placeins}

\usepackage{placeins}

\usepackage{tikz}

\usepackage[framemethod=tikz]{mdframed}

\usepackage{afterpage}

\usepackage{stfloats}

\usepackage{atbegshi}
\newcommand{\handlethispage}{}
\newcommand{\discardpagesfromhere}{\let\handlethispage\AtBeginShipoutDiscard}
\newcommand{\keeppagesfromhere}{\let\handlethispage\relax}
\AtBeginShipout{\handlethispage}

\usepackage{comment}

\title{\LARGE \bf
A Multi-Variate Triple-Regression Forecasting Algorithm for Long-Term Customized Allergy Season Prediction}

\author{Xiaoyu Wu$^{1}$, Zeyu Bai$^{2}$, Jianguo Jia$^{3}$ and Youzhi Liang$^{4}$
\thanks{$^{1}$Boston University, Boston MA USA,
Email: xywu@bu.edu}%
\thanks{$^{2}$University of California Los Angeles, Los Angeles CA USA, Email: zeyubai21@engineering.ucla.edu}%
\thanks{$^{3}$Deakin University, Victoria Australia, Email: j.jianguo@gmail.com}%
\thanks{$^{4}$Massachusetts Institute of Technology, Cambridge MA USA, Email: youzhil@mit.edu}%
}

\begin{document}
\maketitle
\thispagestyle{empty}
\pagestyle{empty}

\begin{abstract}
In this paper, we propose a novel multi-variate algorithm using a triple-regression methodology to predict the airborne-pollen allergy season that can be customized for each patient in the long term. To improve the prediction accuracy, we first perform a pre-processing to integrate the historical data of pollen concentration and various inferential signals from other covariates such as the meteorological data. We then propose a novel algorithm which encompasses three-stage regressions: in Stage 1, a regression model to predict the start/end date of a airborne-pollen allergy season is trained from a feature matrix extracted from 12 time series of the covariates with a rolling window; in Stage 2, a regression model to predict the corresponding uncertainty is trained based on the feature matrix and the prediction result from Stage 1; in Stage 3, a weighted linear regression model is built upon prediction results from Stage 1 and 2. It is observed and proved that Stage 3 contributes to the improved forecasting accuracy and the reduced uncertainty of the multi-variate triple-regression algorithm. Based on different allergy sensitivity level, the triggering concentration of the pollen – the definition of the allergy season can be customized individually. In our backtesting, a mean absolute error (MAE) of 4.7 days was achieved using the algorithm. We conclude that this algorithm could be applicable in both generic and long-term forecasting problems. 

\end{abstract}

\section{Introduction}

Airborne-pollen allergy is prevalent, affecting up to 40\% of the total population worldwide~\cite{d2007allergenic,sanchez2002use}. The long-term customized forecasting of pollen allergy 
provides individuals with guidance for travel planning and medication planning~\cite{prank2013operational,arizmendi1993time,ranzi2003forecasting}. For pharmaceutical companies, the demand of medication for pollen allergy treatment, in addition to the sales and operations planning for supply chain management, further necessities the forecasting of the start/end date of the allergy season~\cite{singh2006supply,jaipuria2014improved,barlas2011demand}. 

\begin{figure}[h]
\centering
\includegraphics[width=0.85\columnwidth]{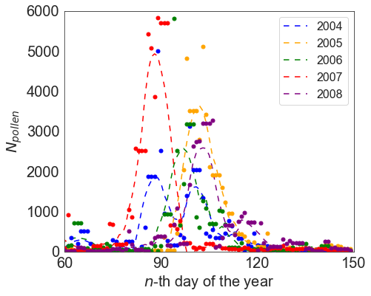}
\caption{The concentration of pollen as a function of the $n$-th day of the year within a 120-day range, i.e. the count of pollen in a cubic meter of air, measured per day for the year from 2004 to 2008. The dashed lines indicate smoothed trend using the Savitzky–Golay filter, merely serving for the visualization.}
\label{Fig-Intro_Pollen}
\end{figure}

In Fig.~\ref{Fig-Intro_Pollen}, the concentration of pollen across the years from 2004 to 2008 is shown. We observe a significant change of the start date and end date for varying years with no explicit trend. For example, the allergy seasons in 2005 and in 2007 exhibit almost no overlapping; the length of the allergy season in 2004 roughly doubles that in any other year. 

For a more rigorous definition of the allergy season tailored for each patient, we introduce the concentration threshold $\delta_C$ which is the minimum customized concentration requirement of pollen for a typical day in the allergy season, and the number threshold $\delta_N$ which is the minimum number of typical days within a week (7 consecutive days) in the allergy season. The start date of allergy season is defined as the day, during the following week of which the number of typical days (pollen concentration $> \delta_C$) is at least $\delta_N$. Both $\delta_C$ and $\delta_N$ can be customized according to different allergy sensitivity level of individuals. For example, if $\delta_C=120$ and $\delta_N=4$, the standard deviation of the start date, the end date and the length of allergy season are 19.9 days, 41.4 days and 47.2 days, respectively, from the year 2003 to year 2019. 

Univariate forecasting methodologies, such as Exponential Smoothing, ARIMA and T-BATS, can achieve expected performance on time series data with a strong signal of level, trend and seasonality~\cite{gardner1985exponential,hillmer1982arima,hassani2017forecasting}. In addition to the classic univariate models, the use of network and neural network architectures, such as convolutional neural network (CNN) and recurrent neural network (RNN) furthers the scope of application~\cite{bai2019mental,chavez2019identify}. Ordinary/Partial differential equations used widely in the field of physics and robotics shed light on the machine learning models, further driving the univariate forecasting investigation recently~\cite{chen2011time,oreshkin2019n, liang2019solid, fish2019dynamic, liang2018dynamic}.

However, univariate forecasting methodologies exhibit underperformance in the context of cyclic intermittency, in particular for predicting the start date and the end date of the allergy season~\cite{stock1998comparison}. In the scenario of airborne-pollen allergy season prediction, the concentration of pollen, primarily produced by plants, is closely dependent on local environmental conditions like the weather and geography, necessitating the integration of the meteorological information into the forecasting methodology~\cite{samal2019time,meese1984comparison}. Univariate models, such as Holt-Winters exponential smoothing, ARIMA and Facebook Prophet model, are unable to integrate other time-varying covariates, in particular the weather information such as the precipitation, temperature and wind~\cite{fildes1984choice,de2011forecasting}. 

In this paper, we propose a multi-variate triple-regression algorithm to predict the airborne-pollen allergy season in the long term, i.e. the start date and end date of the season. The triggering concentration of the pollen – the definition of the allergy season can be customized as aforementioned. The proposed algorithm leverages the inferential signal from other covariates to make long-term accurate predictions, and uses a novel three-stage modelling approach to improve forecasting performance. In particular, we take into consideration the other 11 covariates including the history of temperature, wind and precipitation in addition to the historical data of pollen concentrations. The prediction results from early stage(s) are used in later regressions to further improve the accuracy and reduce the uncertainty of prediction. 

\section{Algorithm}
\begin{figure}[h]
\centering
\includegraphics[width=1\columnwidth]{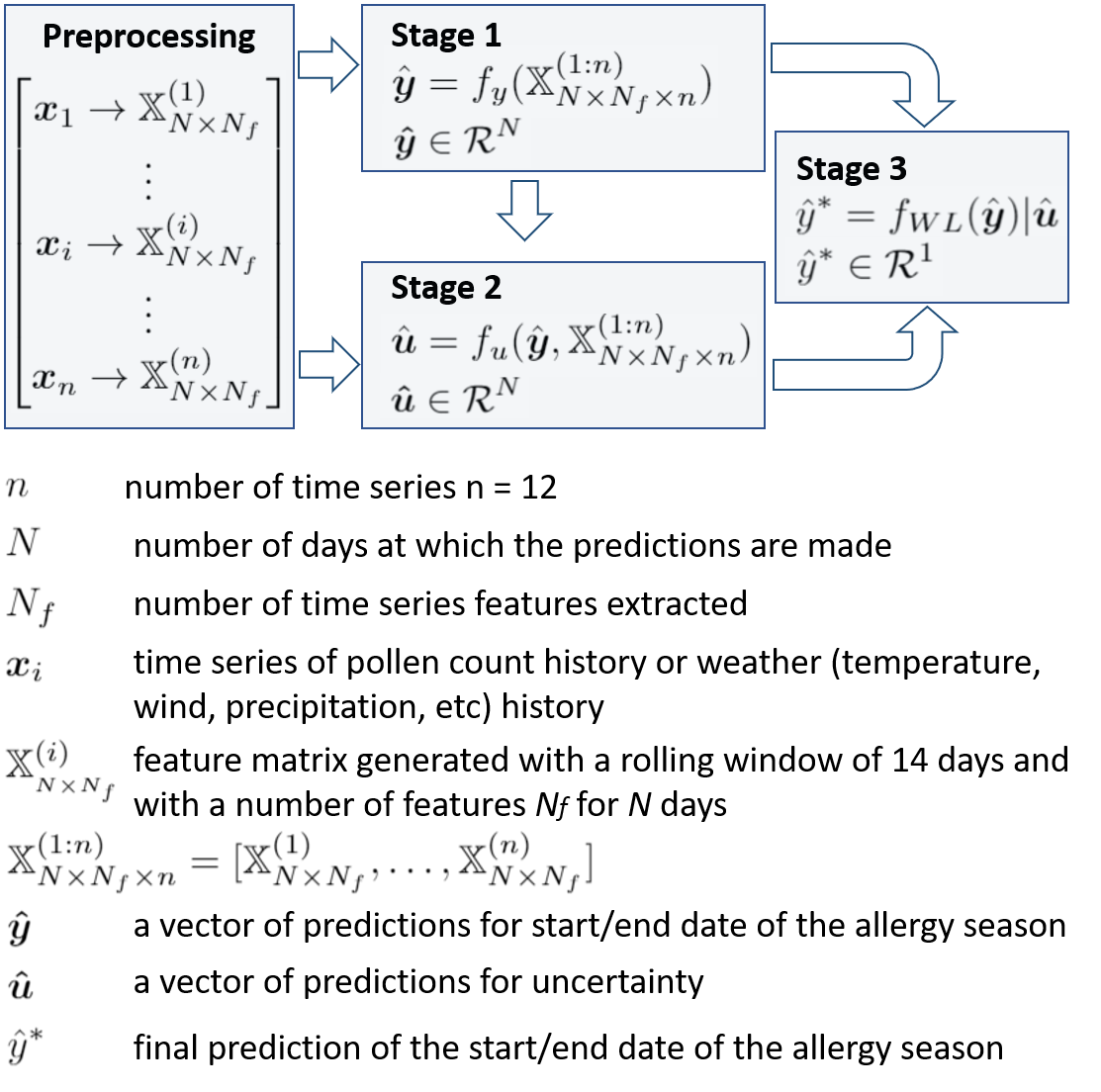}
\caption{Data pipeline of the proposed triple-regression algorithm with a nomenclature. The feature matrix from pre-processing is the input for Stage 1; the outcome from pre-processing and Stage 1 is the input for Stage 2; the regression results from Stage 1 and Stage 2 are used in Stage 3.}
\label{Fig-Alg-flow}
\end{figure}

The proposed algorithm encompasses a three-stage regression for the start/end date of the allergy season prediction, together with a pre-processing for the feature extraction. The data pipeline and the regression algorithm are outlined in Fig.~\ref{Fig-Alg-flow}. In the pre-processing, a total of ${N_f}$ (=30) time series selected features are extracted by applying a 14-day rolling window to each of the ${n}$ (=12) original time series vectors $\boldsymbol{x}_i$, including the pollen concentration history, temperature history, wind history, participation history, etc. The feature matrix corresponding to $\boldsymbol{x}_i$ is denoted as $\mathbb{X}^{(i)}_{N \times N_f}$, and the ensemble feature matrix is $\mathbb{X}^{(1:n)}_{N \times N_f \times n}$. The ensemble feature matrix is then fed into the three-stage regression. In Stage 1, a Gradient Boosting model to predict the start/end date is trained on a training set $\mathbb{S}_1$ which is based on the feature matrix $\mathbb{X}^{(1:n)}_{N \times N_f \times n}$. The ground truth for the start/end date of allergy season is determined following the definition after we select appropriate pollen concentration threshold $\delta_C$ and number threshold $\delta_N$. The vector of prediction is given by $\hat{\boldsymbol{y}} = f_{y}(\mathbb{X}^{(1:n)}_{N \times N_f \times n}) | \mathbb{S}_1$. In Stage 2, we select another training set $\mathbb{S}_2$ based on the feature matrix $\mathbb{X}^{(1:n)}_{N \times N_f \times n}$ and the predicted start/end date $\boldsymbol{\hat{y}}$ from Stage 1 to train another Gradient Boosting model to predict the uncertainty in $\boldsymbol{\hat{y}}$. The vector of uncertainty is given by $\hat{\boldsymbol{u}}=f_u(\hat{\boldsymbol{y}},\mathbb{X}^{(1:n)}_{N \times N_f \times n}))|\mathbb{S}_2$. In Stage 3, we perform a weighted linear regression on $\boldsymbol{\hat{y}}$ based on the linear constraint on start/end date predictions at consecutive days ahead of the allergy season. The weights are assigned according to the predicted uncertainty $\boldsymbol{\hat{u}}$. Thus, the final predicted start/end date of the allergy season is obtained by $\hat{y}^* = f_{WL}(\hat{\boldsymbol{y}}) \vert \boldsymbol{W}=\hat{\boldsymbol{u}}$.

Although one can opt to terminate the algorithm at Stage 1 when the predictions are already made, we emphasize that the three-stage regression can guarantee its uncertainty to be smaller than that using only one-stage regression. The proof is given for a simplified problem as follows. 

We may assume that each prediction in the first-stage regression $\hat{y}_i$ follows a Gaussian distribution:
\begin{equation}
    \hat{y}_i \in \mathcal{N}(\mu_i, \sigma_i^2),
\end{equation}
where the variance $\sigma_i$ is assumed to be constant $\sigma_0$, and $\mu_i$ is theoretically constrained by a linear relationship given the predictions are made at consecutive days $z_i$. To set up the problem of a multi-linear regression in general, we may consider $p$ independent variables $x_1, ..., x_n$ and one dependent variable $y$. Suppose we have $n \, (n>p)$ observations,
\begin{equation}
    \hat{y}_i = \beta_0 + \sum \beta_i z_{ij} + \epsilon_i, \, i = 1,...,n,
\end{equation}
where $\beta_i$ are the coefficients of the $i$-th dependent variable $x_i$. Our goal is to minimize the sum of the weighted squared residuals (errors) $\epsilon_i$. Thus, the cost function is:
\begin{equation}
    \sum _{i=1} ^n w_i \epsilon_i^2 = \sum _{i=1} ^n w_i \left( \hat{y}_i - \beta_0 - \sum _{j=1} ^p \beta_j z_{ij} \right)^2,
\end{equation} 
where the weights $w_i$ is provided by the inverse of the uncertainty $\hat{u}_i^2 |_{(2)}$ from the regression results in Stage 2, $w_i = 1/\sigma_i^2 |_{(2)} = 1/\hat{u}_i^2 |_{(2)}$. No analytical solution can be obtained for a set of random weights. Without loss of generality, we can focus on a simplified scenario with uniform weights, and the linear regression results can be expressed as:
\begin{equation}
    \boldsymbol{\hat{Y}} = \mathbb{E}[\boldsymbol{Y}] + \frac{\mathbf{Cov}(\boldsymbol{Z},\boldsymbol{Y})}{\mathbf{Var}(\boldsymbol{Z})} (\boldsymbol{Z}-\mathbb{E}[\boldsymbol{Z}]),
\end{equation}
where $\mathbb{E}[\boldsymbol{Y}]$ is the expectation of random variable $\boldsymbol{Y}$, $\mathbf{Cov}(\boldsymbol{Z},\boldsymbol{Y})$ is the covariance of random variable $\boldsymbol{Z}$ and $\boldsymbol{Y}$, and $\mathbf{Var}(\boldsymbol{Z})$ is the variance of random variable $\boldsymbol{Z}$.

\begin{figure}[h]
\centering
\includegraphics[width=0.9\columnwidth]{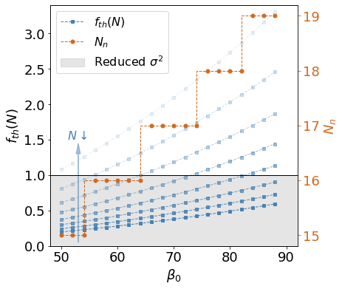}
\caption{Threshold function $f_{th}(N)$, and minimum number of days $N_n$ to reduce uncertainty, as functions of the regression coefficient $\beta_0$ with varying number of days $N$ used for prediction. The shaded area indicates the regime where the variance using a three-stage regression is reduced compared with the one-stage regression counterpart. Note that $\beta_1$ is assumed to be 1 under the condition of ideal scenario.}
\label{Fig-Alg-theory}
\end{figure}

\begin{figure}[h]
\centering
\includegraphics[width=0.9\columnwidth]{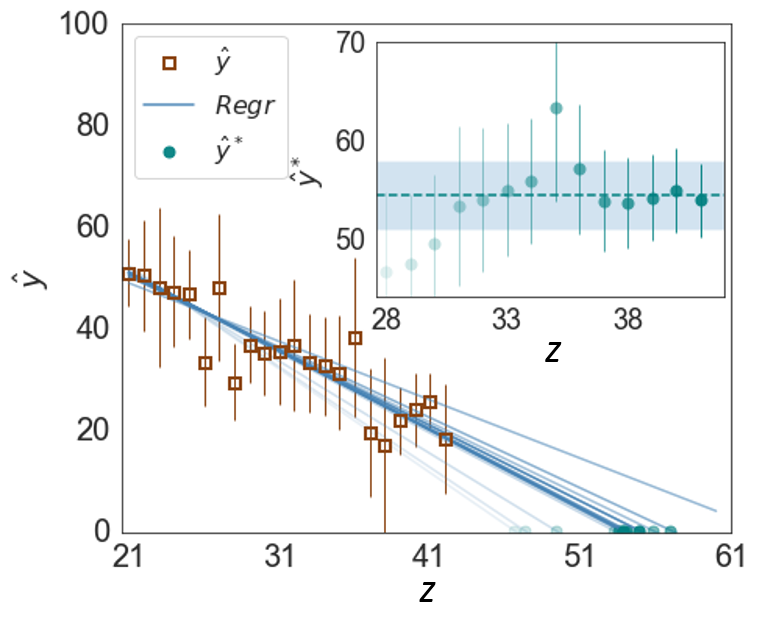}
\caption{Start date $\hat{y}$ predicted in Stage 1, as a function of the $z$-th day of the year 2005, where the error bar indicates the standard deviation predicted in Stage 2. Straight lines are the weighted linear regression results with different number of  predictions considered. (Inset) Final prediction of the start date $\hat{y}^*$ predicted in Stage 3, as a function of the $z$-th day of the year. Shaded area indicates the band of standard deviation.}
\label{Fig-Discussion-DoubleStageRegr}
\end{figure}

In the context of predicting the start/end date of the allergy season, the only independent variable in the weighted linear regression is the date $z_i$ at which the prediction is made. Thus, we only have two non-zero coefficients, $\beta_0$ and $\beta_1$, to be learned from the regression. The variance of the predicted coefficient can be approximated by~\cite{james2013introduction}:
\begin{equation}
    \begin{split}
        \sigma^2(\hat{\beta}_0) &= \sigma'^2 \left( \frac{1}{n} + \frac{\overline{z}^2}{\sum\limits_0^N (z_i-\overline{z})^2} \right),\\
        \sigma^2(\hat{\beta}_1) &= \sigma'^2 \frac{1}{\sum\limits_0^N (z_i-\overline{z})^2},
    \end{split}
\end{equation}
It can be shown that $\sigma'^2$ is related to the uncertainty from the Gradient Boosting regression in Stage 1 through the following formula:
\begin{equation}
    \sigma'^2 = \frac{\sigma_0^2}{N}.
\end{equation}
The final prediction of the start/date date $\hat{y}^*$ is given by the $z$-intercept of the line from the weighted linear regression:
\begin{equation}
    \begin{split}
        \hat{y}^* = - \frac{\beta_0}{\beta_1}
    \end{split}
\end{equation}
Thus, the standard deviation of $\hat{y}^*$ is approximated by
\begin{equation}
    \begin{split}
        &\sigma(\hat{y}^*) = \hat{y}^* \sqrt{\frac{\sigma'^2(\hat{\beta}_0)}{\hat{\beta}_0^2}+\frac{\sigma'^2(\hat{\beta}_1)}{\hat{\beta}_1^2}}\\
        &= - \frac{\hat{\beta}_0}{\hat{\beta}_1} \sqrt{ \frac{\sigma'^2}{\hat{\beta}_0^2} \left( \frac{1}{n} + \frac{\overline{x}^2}{\sum (x_i-\overline{x})^2} \right) + \frac{\sigma'^2}{\hat{\beta}_1^2} \frac{1}{\sum (x_i-\overline{x})^2} }
    \end{split}
\end{equation}
Utilizing the third-stage regression, we aim for a reduced variance, i.e. $\sigma^2(\hat{y}^*) < \sigma^2_0$. In other words, the uncertainty should be reduced from the one-stage regression result. It is obvious that a minimum number of days $N_{n}$ where the predictions are made in Stage 1 is required to guarantee the reduced uncertainty. We can calculate $N_{n}$ by first defining the threshold function:

\begin{equation}
    \begin{split}
        f_{th}(N_{n}) =& \frac{1}{N_n\hat{\beta}_1^2} \left( \frac{1}{N_{n}} + \frac{\overline{x}^2}{\sum\limits_0^{N_{n}} (x_i-\overline{x})^2}  +  \frac{{\hat{\beta}_0^2}/{\hat{\beta}_1^2}}{\sum\limits_0^{N_n} (x_i-\overline{x})^2} \right),
    \end{split}
\end{equation}
then setting $f_{th}(N_{n})$ to 1, indicating that the uncertainty does not change after the weighted linear regression.

\section{Results and Discussion}

To guarantee a reduced uncertainty in the three-stage regression compared with the one-stage regression, we obtain the minimum number of days, $N_n$, used for making predictions. In Fig.~\ref{Fig-Alg-theory}, we plot the threshold function $f_{th}(N)$ as a function of the coefficient $\beta_0$ with varying $N$. The shaded area in Fig.~\ref{Fig-Alg-theory} indicates the regime where the three-stage regression has a reduced uncertainty. 

Therefore, the corresponding minimum number of days $N_n$ for a specific $\beta_0$ value can be obtained from the threshold function which satisfies $f_{th}(N) < 1$ and has the smallest $N$ among all threshold functions. As $\beta_0$ increases, the value of threshold function increases, reflecting an increased minimum number of days $N_n$ is required. 

We apply the data pipeline and three-stage regression algorithm to the dataset described in Section II to predict the start date of the allergy season. The ground truth of the start date of the allergy season is set by the thresholds,  $\delta_C=120$ and $\delta_N=4$, i.e., for seven consecutive days after the start day, the number of days when the pollen concentrations is greater than $\delta_C$ is at least $\delta_N$. 

In Fig.~\ref{Fig-Discussion-DoubleStageRegr}, we show the mean $\hat{y}$ predicted in Stage 1, and the corresponding error bar, which is the standard deviation $\sigma(\hat{y})$ predicted in Stage 2 as functions of the $z$-th day of the year 2005. The models to predict $\hat{y}$ and $\sigma(\hat{y})$ are trained using the dataset from year 2003 and 2004 respectively. The blue lines in decreasing transparency represent weighed linear regression results with increasing number of predictions $\hat{y}$ used, where the inverse of the uncertainty $\sigma(\hat{y})$ serves as the weights. The green circles locates at the $z$-intercept represents the final prediction made in Stage 3. It is manifested in the Fig.~\ref{Fig-Discussion-DoubleStageRegr} (Inset) that the prediction converges to Day 54 as the number of days accounted increases, while the actual start date is Day 51 for year 2005. We also performed backtesting for year 2006, 2007 and 2008, and a mean absolute error of 4.7 days was achieved using the triple-regression algorithm.

The intercept of $y$-axis, coefficient $\beta_0$, indicates the approximate prediction of the start/end date.

\section{Conclusions}

The airborne-pollen allergy season exhibits significant variations in terms of the start/end date and the length of the allergy season. Univariate models fail to extract its seasonality or trend and fail to integrate other covariates such as the temperature and precipitation. 

In our proposed triple-regression algorithm, (a) the pollen allergy season can be customized based on each individual's allergy sensitivity level, (b) the predictions are obtained based on the historical data of pollen concentration together with other 11 covariates, (c) most importantly, the uncertainty of the prediction in Stage 3 can be reduced given that the minimum number of predictions obtained from Stage 1 is satisfied. The final prediction in Stage 3 converges to a mean value with the increasing number of predictions obtained from Stage 1. The weighted linear regression further improve the accuracy by integrating the uncertainty predicted in Stage 2. In our backtesting, a mean absolute error of 4.7 days was achieved using the triple-regression forecasting algorithm. We conclude that this algorithm could be useful in both generic and long-term forecasting problems.


\bibliographystyle{ieeetr}
\bibliography{bibliography}

\clearpage
\end{document}